\documentclass[amstex,aps,amsfonts,prsty]{article}

\usepackage{amssymb}
\usepackage{amsmath}
\usepackage{graphicx}
\usepackage{citesort}

\def \be{\begin{equation}}
\def \ee{\end{equation}}
\def \ba{\begin{array}{l}}
\def \Ba{\begin{array}{ll}}
\def \ea{\end{array}}
\def \bq{\begin{eqnarray}}
\def \eq{\end{eqnarray}}
\def \nn{\nonumber\\}
\def \lb{\label}

\def \fr{\frac}
\def \a{\alpha}
\def \b{\beta}
\def \d{\delta}
\def \D{\Delta}

\def \f{\phi}

\def \lm{\lambda}

\def \n{\nabla}
\def \p{\varphi}

\def \s{\sigma}
\def \t{\tau}

\def \tl{\tilde}
\def \ol{\overline}

\def \la{\langle}
\def \ra{\rangle}
\def \[{\left[}
\def \]{\right]}
\def \({\left(}
\def \){\right)}

\def \T{T_{c}}

\def \I{\int d^{D}x}
\def \2{\frac{1}{2}}
\def \4{\frac{1}{4}}

\begin{document}

\begin{center}

{\Large \bf On the nature of the phase transition in the three-dimensional
            random field Ising model}

\vskip .2in

Vik.S.\ Dotsenko

\vskip .1in

LPTMC, Universit\'e Paris VI,  4 place Jussieu, 75252 Paris, France  

L.D.Landau Institute for Theoretical Physics, 
   117940 Moscow, Russia

\end{center}

\vskip .3in

\begin{abstract}
A brief survey of the theoretical, numerical and experimental studies of the
random field Ising model  during last three decades is given.
Nature of the phase transition in the three-dimensional  RFIM with Gaussian
random fields is discussed. Using simple scaling arguments it is shown that 
if the strength of the random fields is not too small (bigger than a certain threshold value) 
the finite temperature phase transition in this system is equivalent to the low-temperature 
order-disorder transition which takes place at variations of the strength of the random fields.
Detailed study of the zero-temperature phase transition 
in terms of simple probabilistic arguments and modified mean-field approach
(which take into account nearest-neighbors spin-spin correlations) is given.
It is shown that if all thermally activated processes are suppressed the ferromagnetic
order parameter $m(h)$  as the function of the strength $h$ of the random fields becomes
history dependent. In particular, the behavior of the magnetization curves $m(h)$ for
increasing and for decreasing $h$ reveals the hysteresis loop.
\end{abstract}

\vskip .1in


\vskip .3in

\section{Introduction}

The thermodynamic properties of the random field Ising model (RFIM)  remains 
the controversial issue during more that 30 years now 
(for reviews see e.g. \cite{rev,rev-exp}). Traditionally there are exist three somewhat
independent mainstreams of reseach: theoretical, numerical and experimental.
Inside each of them,  as well as between any of them, 
one finds broad divergences in the obtained results and in their interpretations.
The main issue of this controversy is the structure of the phase diagram
of the system: what kind of phase transitions and which thermodynamic phases are present there?

The random field model has been proposed originally by Larkin \cite{larkin}
to model the defect pining of vortices in sumerconductors.
The most simple version of the this model for systems with discrete
Ising symmetry, the random field Ising model (RFIM), can be
described in terms of the Hamiltonian

\be
\label{rfim0}
H \; = \; - \sum_{<i,j>}^{N} \sigma_{i} \sigma_{j} \; - \; \sum_{i} h_{i} \sigma_{i}
\ee
where the Ising spins $\{ \sigma_{i} = \pm 1 \}$ are placed in the vertices
of a D-dimensional lattice with the ferromagnetic interactions between
the nearest neighbors, and quenched random fields $\{ h_{i} \}$ are
described either by the symmetric Gaussian distribution with
$\la h_{i} h_{j}\ra = \d_{ij} h^{2}$, or by the binary distribution
in which the fields are taking two values $\pm h$ with independent probabilities
at each site.

%%%%% THEORY
\subsection{Theory}

The first round of controversies in the studies of RFIM 
was about its lower critical dimension.
According to simple energy ballace physical arguments by Imry and Ma \cite{Imry-Ma}
it should be expected that the dimensions $D_{c}$ above which the ferromagnetic
ground state is stable at low temperatures (it is called
the lower critical dimension) must be $D_{c} = 2$ (unlike the pure Ising
systems where $D_{c} = 1$).
  Later, the existence of the ferromagnetic long range order
in the 3D RFIM has been confirmed by a rigorous proof \cite{rigorous}.
  On the other hand, a perturbative renormalization group study of 
the phase transition demonstrates the so called phenomenon of a dimensional 
reduction, such that the critical exponents of the random field system 
in the dimension D appear to be the same as those of the ferromagnetic 
system without random fields in the dimension d=D-2 \cite{D-2},
which  implies that the lower critical 
dimension of the random field Ising model must be equal to 3, in
contradiction with the rigorous results. Actually, the procedure of summation
of the leading large scale divergences assumes that the Hamiltonian 
has only one minimum. 
However, one can easily see that as soon as the temperature is close enough
to the putative critical point (as well as in the whole low temperature 
region), there are local values of the
magnetic fields for which the free energy has more than one minimum
\cite{many-solu-pari,manu-solu-numer}, and  therefore the
dimensional reduction is not grounded. Thus, the above arguments have settled the
controversy about the lower critical dimension of the random field Ising
model in favor of the value $D_{c} = 2$.

Phase diagram in the (h,T) plane for the RF Ising system has been first
derived in terms of simple mean-field theory for the model with infinite-range 
ferromagnetic interactions (which corresponds to infinite space dimensionality) 
with the Gaussian distribution of random fields \cite{mfG}. 
In this system the ferromagnetic phase is separated 
from the paramagnetic phase by a line $T(h)$ of the second order phase transition which 
monotonically decreases with $h$ from the point $T_{c}(0)$ of the pure system,
down to the ending point $T=0$ at $h=h_{c}$.
  On the other hand, for the bimodal distribution \cite{mfB} one finds that 
the line $T_{c}(h)$ contains a tricritical point $(T_{*}, h_{*})$,
such that at $h > h_{*}$ the phase transition becomes the first order.
  Later on a simple scaling description of the RFIM phase transition in finite
dimensions has been developed assuming that it is of the second order, and that 
it is controlled by the zero-temperature fixed point \cite{scaling}.

As another extreme case, instead of the Ising spins,  one can study the $m$-component 
vector  spin system (the Ising model corresponds to $m=2$) with Gaussian random fields.
It can be shown \cite{rf-rsb} that in terms of the replica field theory
in the limit $m \to \infty$ the statistics of interfaces in this system is described by 
the solution with broken replica symmetry (similar to that in spin glasses \cite{sg-rsb}) 
which is consistent with the low critical dimension $d_{l} = 2$.
Later on in the same limit $m \to \infty$ it has been demonstrated \cite{rf-rsb-Mez-Young}
the instability of the replica symmetric state and the onset of the replica
symmetry breaking scenario. In terms of this approach it has been argued that
in the phase diagram of the three-dimensional RFIM the the paramagnetic and ferromagnetic
phases are separated by the glassy phase in which the replica symmetry is broken \cite{glassy-mez-monas}.

An independent study of the $\phi^{4}$ replica field theory (which in the critical
region is believed to describe RFIM) in terms of the Legendre transforms technique
and the virial-like relations also shows the presence of the intermediate RSB glassy
phase below dimensions $D=6$ \cite{glassy-domini}.
Further renormalization-group (RG) study of this field theory in dimensions $D < 6$
revealed no stable fixed points signalling the nonset of the replica symmetry breaking
\cite{rsb-domini}. 

Another line of theoretical studies concerns the non-perturbative thermodynamic states
which are apparently missing in the usual RG calculations and 
which may appear to be quite relevant for the nature of the phase transition in RFIM.
The importance of the non-perturbative phenomena has been noted already long time
ago \cite{many-solu-pari,pari-dos,rf-grif-dos}. In physical terms, the non-perturbative
configurations are rare spatial regions with "flipped" (opposite to the background)
magnetizations, and in terms of the replica field theory they are described by
localized in space finite energy instanton configurations in which the replica symmetry
is broken. In the recent investigations
a systematic approach for the calculation of the non-perturbative contributions 
has been developed \cite{states-dos} and it has been shown that 
from the point of view of their relevance the dimension $D=3$  turns out to be marginal: 
formally these degrees of freedom produce finite (non-analytic 
in the strength of the field $h$) contribution only in dimensions $D \leq 3$
\cite{3d-non-pert}.

Finally, the construction of an alternative approach, the functional renormalization 
group, which is supposed to take into
account all non-perturbative degrees of freedom for the whole class 
of the random-field $O(N)$ spin model is under way these very days \cite{func-rg}

\subsection{Numerics}

%%%%%%%% NUMERICS

Numerical investigations of the phase transition in RFIM 
reveal broad controversy in the obtained results and in particular
in their interpretation. The main point of the disagreements is about  the nature
if the phase transition: is it continuous, or is it of first order with finite jump
of the order parameter.
 
Originally, the analysis of the high-temperature series expansion 
for the RFIM with the Gaussian distribution of random fields \cite{high-T-series-A}
indicated on the existence of a fluctuation-driven first order phase transition
at sufficiently strong disorder below four dimensions (which suggests the existence 
of a tricritical point in the three-dimensional system). However more recent study
indicated on the continuous transition, at leas for weak disorder \cite{high-T-series-B}.

First extensive Monte Carlo (MC) numerical simulations for the 3D RFIM have been carried out by
Young and Nauenberg \cite{MC-1-order}, who reported that the transition may be first order.
On the other hand, the numerical study by Ogielski and Huse \cite{MC-2-order-A},
and by Ogielski \cite{MC-2-order-B} shows that the transition may be second order.
More recent Monte Carlo simulations \cite{MC-2/1} can also been interpreted as
describing a continuous transition but with a finite jump in the magnetization.
However, later extensive MC  calculations \cite{MC-1-order-B} for the energy and 
the magnetization distributions suggest a first order transition. 

The zero-temperature studies play an important role in understanding the nature of the
phase transition in the RFIM. According to the zero temperature fixed point 
hypothesis, the transitions which take place at $T=0$ and at $T\not= 0$ are in the same
universality class. First zero-temperature 
Migdal-Kadanoff renormalization-group calculations indicated a continuous transition
which is characterized by very small value of the  magnetization critical exponent
($\beta \simeq 0.02$) \cite{fin-size-2-T=0}. Similar results were obtained
in the finite temperature renormalization group study \cite{fin-size-2-T}.
On the other hand, some of more recent zero-temperature finite size scaling 
calculations \cite{fin-size-1} have demonstrated the existence of a first order phase transition,
although most of the them \cite{fin-size-2-B,fin-size-2-C}
are in favor of the continuous transition. Moreover, in one of the most recent calculations
it has been demonstrated that for a given realization of disorder the ground states ($T=0$)
and thermal states ($T\not= 0$) near the critical line are strongly correlated, which may be
interpreted as the concrete manifestation of the zero temperature fixed point scenario 
\cite{fin-size-2-T=0-T}. 
The most precise results obtained by 
A.A.Middleton and D.S.Fisher \cite{fin-size-2-C} are clearly consistent with the 
existence of a single phase transition (no intermediate spin-glass phase has been observed)
and it is characterized by very small value of the magnetization critical exponent
$\beta = 0.017$. 

Others numerical studies of the phase diagram of the RFIM in the plane $(T, h)$
\cite{phase-dia-A,phase-dia-B}, also find no glassy phase (with or without
replica symmetry breaking) which would separate paramagnetic and ferromagnetic
phases. On the other hand, for the bimodal random field distributions 
a discontinuity in the magnetization at the phase transition 
line is observed \cite{phase-dia-B}.

Finally, the numerical simulations of the diluted antiferromagnet in a field
(DAFF) designed to imitate the random field magnetic systems which are
studied  experimentally (see below), shows that in equilibrium, the transition to the LRO 
state may be first order \cite{daff-1,daff-2}, and moreover it is claimed that at sufficiently strong
dilution there is first order phase transition to a spin glass state \cite{daff-sg}

To conclude this brief survey of numerical studies, it should be noted
that all numerical simulations in RFIM are impeded by the dramatic slowing down when
approaching a putative phase transition, which very often makes very difficult
to interpret the obtained results. Besides, the delicate point in all these studies 
is that the order parameter critical exponent $\beta$ turns out to be very small, 
so it is very difficult to distinguish
the second order transition (for which the order parameter continuously goes to zero
at $T_{c}$) from the first order (when the order parameter has a finite jump at $T_{c}$). 
   On top of that, recent numerical study of the critical region of the 3D RFIM 
indicates that the situation could be even more sophisticated due to absence 
of self-averaging of the correlation length, of the specific heat, and maybe of some others 
thermodynamical quantities \cite{non-self-ave}.

\subsection{Experiments}

%%%%% EXPERIMENTAL

From the point of view of experimental investigations
an important theoretical breakthrough occurred when it has been realized that
RFIM can be generated in diluted Ising antiferromagnets by application
of a {\it uniform} external magnetic field \cite{rf-realiz}. 
This has opened the way for experimental studies of RFIM, 
and in particular for the investigation of the phase transition.

In the very first experimental studies of the RFIM performed in the diluted 
antiferromagnets $Co_{x}Zn_{1-x}F_{2}$ it has been claimed that even the smallest
magnetic fields destroy the long-range order at all temperatures 
in three dimensions \cite{exper-no-LRO}.

First convincing experimental demonstrations of the existence of the phase transition 
in the 3D random field antiferromagnet $Fe_{x}Zn_{1-x}F_{2}$  and its critical properties 
(assuming that it is of second order) has been reported in \cite{exper-2A}.
The extensive study of the critical properties of the system $Fe_{0.6}Zn_{0.4}F_{2}$
for temperatures $T > T_{c}(h)$ has been done in \cite{exper-2B}.

In fact, the behavior observed near the critical point depends on the field and 
temperature procedures used in the measurements. In the field cooled (FC) samples
the transition appears to be rounded, and at low temperatures the system 
freezes in the metastable domain state. However, if the sample is cooled in 
the zero field (ZFC), the transition appears to be much sharper,
and rounded only by slow dynamic effects \cite{rev-exp}.

On the other hand, in another experimental study of the 3D RFIM (for weakly anisotropic 
$Mn_{0.5}Zn_{0.5}F_{2}$) \cite{exper-1A} it has been observed that when approaching the
phase transition the correlation length reaches a finite size, at which point
it was assumed that further approach towards a putative second order transition 
is interrumpted by occurrence of a first order transition.

The most slippery issue of the experimental studies of the phase transition in the RFIM
is the measurement of the static critical behavior of the staggered magnetization.
Similarly to the numerical studies, the fact that critical exponent $\beta$
is experimentally very small (e.g. according to \cite{exper-beta1} $\beta \simeq 1/8$ or less,
according to \cite{exper-beta2} $\beta \simeq 0.16$)
makes it very difficult to work out any definite conclusion
about the nature of the observed phase transition.

In the studies of the diluted antiferromagnet $Mn_{0.75}Zn_{0.25}F_{2}$ the 
field-cooled data were interpreted as the evidence for an equilibrium 
second-order transition at$T_{FC}$, while the zero-field cooled data 
were interpreted as non-equilibrium {\it trompe l'oeil} transition
in which the long-range order diminishes continuously at $T_{ZFC} > T_{FC}$, 
while the correlation length only reaches a finite maximum value
at a temperature between $T_{ZFC}$ and $T_{FC}$ \cite{expere-FC-ZFC-A}.
Similar behavior was observed in $Fe_{0.5}Zn_{0.5}F_{2}$ \cite{expere-FC-ZFC-B}.
On the other hand, all these results can be explained as the mean-field first-order transition
broadened by the cluster-flipping mechanism \cite{expere-1-order-distr}.

In the study of the random field antiferromagnet $Fe_{0.5}Zn_{0.5}F_{2}$
a notable broadening of the phase transition region has been observed, so that
the data may be described by the Gaussian distribution of effective transition temperatures
with the width which scales with the strength of the applied field as $h^{2}$ 
\cite{exper-Tc-distr}. Besides, the apparent critical behavior observed in these
measuremets represented a continuous evolution from metastable behavior
towards equilibrium behavior. Moreover, according to the point of view of Birgeneau
\cite{exper-no-equilib}, so far no true equilibrium phase transition has been observed
in RFIM systems. Nevertheless, in the recent study of high-magnetic concentration
Ising antiferromagnet 
$Fe_{0.93}Zn_{0.07}F_{2}$ (which does not exhibit the severe critical scattering
hysteresis \cite{exper-high-mag-concent})
it is claimed that in terms of general scaling assumptions
the critical scattering analysis allows to reveal equilibrium critical behavior
and to obtain the corresponding critical exponents \cite{exper-crit-exp} 
(in particular the specific heat critical behavior is well fitted by the logarithmic 
singularity which corresponds to $\alpha = 0$ \cite{exper-crit-alpha=0}).

\vspace{10mm}

In conclusion of this brief historic review one can note that the situation with the 
phase transition in the three-dimensional RFIM remains far from being clear.
Theorists are mostly interested in their own problems (like $1/m$ expansion, or 
RSB in dimensions close to 6, or two-loop approximation for $O(N)$ sigma model 
with the number of spin components close to $N_{c} \simeq 2.835$), 
which are rather far from the original system,
while people doing numerics and in particular
experiments are facing realities full of numerous secondary and accessory effects
which very often overshadow the main physical phenomena.

It is evident that one can not hope to find the exact solution for this problem,
and therefore some kind of simplifications and approximations are unavoidable.
In this paper, which is mostly devoted to the low-temperature part of the phase
diagram,  it will be supposed that: 
{\bf (a)} the finite temperature thermodynamical properties of 
the three-dimensional RFIM can be described in
terms of the continuous Ginzburg-Landau $\phi^{4}$ field theory;
{\bf (b)} we can neglect the presence of randomness in the spin-spin interactions; and
{\bf (c)} in the study of the {\it zero-temperature} phase transition (at variation of the strength
of the random fields) all thermally activated processes are suppressed.

The above reservations are rather essential. 

{\bf (a) Ginzburg-Landau field theory}. We all {\it believe} in the universality,
and therefore if we admit that the phase transition in a given system is of the second order, 
then  it is generally accepted that its critical properties can be described 
in terms of the corresponding Ginzburg-Landau Hamiltonian (with proper symmetry properties). 
However, if the system is considered at a {\it finite} distance from the putative critical point
(which is the case in the present paper) where the correlation length is finite,
then the relevance of the {\it continuous limit} Ginzburg-Landau Hamiltonian for the 
original lattice Ising system becomes much less evident. 

{\bf (b) Randomness in the spin-spin interactions}. 
Apart from the numerical investigations of DAFF \cite{daff-1,daff-sg}, the effects of randomness in
the spin-spin interactions is practically never taken into account in theoretical studies 
(there are enough headache due to random fields...). 
  However, it is well known that the presence of this type of randomness 
(without random fields) may change the critical properties of a magnetic system, and in particular 
it definitely does in the three-dimensional Ising model. Moreover, in some statistical systems
(e.g. in the Potts model) the presence of such disorder may turn the first-order transition 
(of the pure system) into the second order one \cite{first-second}. 
Therefore, it we admit for the moment that the phase transition in the "pure"
(without randomness in spin-spin interactions) 3D RFIM is 
of the first order, it does not guarantee that in the corresponding experimental realizations
of this system (where the randomness in the spin-spin interactions is inevitable) the phase 
transition would not turn into the second order one.

{\bf (c) Thermal activations}. In the studies of the zero-temperature properties of the system
as the function of the effective strength of the random fields
$h$,  one can consider two types of procedures. 
In the first one it is assumed that at any given $h$ the system is at the
ground state (which, at a given realization of the random fields, is assumed to be unique).
In this case the {\it variation}  of $h$ to a new value $h'$ means that either we admit that
the temperature is actually slightly non-zero, and the system passes to a new ground state
configuration (corresponding to $h'$) via thermally activated flips, or at each variation of $h$
the ground state configuration  is achieved by cooling down the system (at fixed $h'$)
from the high-temperature disordered state. 

The alternative procedure of the variation of $h$
is essentially different. Here one has to fix the {\it starting} value of $h$ (e.g. $h=0$), 
and then at any variation of $h$ the spins
are just following  the directions of their  local fields (which consist of the contributions 
due to the regular ferromagnetic interactions with the nearest neighbors 
plus the local value of the random field).
In this procedure all thermal "jumps over energy barriers" are prohibited, and therefore, 
most probably, due to the variation of $h$ the system will get stuck in one of the metastable 
states. 

Of course, from the experimental point of view none of these two procedures 
can be implemented in the pure form. The first one (qualitatively, it corresponds to the 
{\it field cooled} (FC) measurements) requires very long (formally infinite) 
waiting time to reach the thermal equilibrium at any variation of $h$, while the second 
one (it corresponds to the {\it zero field cooled} (ZFC) measurements) requires to suppress {\it all} thermally activated processes which can be achieved only 
if the temperature is exactly equal to zero. 

\vspace{5mm}

 In the next Section using simple physical arguments it will be shown that
if the strength of the random fields is not too small then there exists a region of parameters 
where the critical fluctuations are irrelevant, and therefore 
the thermodynamic properties of the system 
can be described in terms of the saddle-point equations of the corresponding (random) 
Ginzburg-Landau Hamiltonian. It is in this region that the ferromagnetic order breaks down. 
Moreover, using simple scaling arguments it will be shown that in this case
the finite temperature phase transition in this system is equivalent to the low-temperature 
phase transition at variation of the strength of the random fields.

In Section 3 the detailed study of the zero-temperature order-disorder phase transition 
will be done in terms of the discrete Ising model on the three-dimensional cubic lattice.
Here we propose a modified (slightly improved) version of the mean-field approach
which takes into account nearest-neighbors two-spins  correlations. Then,
considering the two regimes: increasing and decreasing strength of the random fields, and
supposing that all thermally activated spin flips are suppressed, it will be demonstrated 
that the magnetization curve $m(h)$  reveals the hysteresis loop. In particular,
it is characterized by two different critical points  $h_{c}^{(-)} < h_{c}^{(+)}$,
where $h_{c}^{(-)}$ is the strength of the random fields at which $m(h)$ becomes
non-zero at {\it decreasing} $h$, and $h_{c}^{(+)}$ is the one at which $m(h)$ becomes
zero at {\it increasing} $h$.

\section{Heuristic arguments}

For the sake of generality, before coming down to the dimension three,
first, let us consider the continuous version of the $D$-dimensional Ising model
in terms of the scalar field Ginzburg-Landau Hamiltonian:
\be
\lb{rfim1}
H\[\f({\bf x}),h({\bf x})\] \; = \; \I \Biggl[ \2 \(\n\f({\bf x})\)^{2}  
   - \2 \t \f^{2}({\bf x})  
   + \4 g \f^{4}({\bf x}) - h({\bf x}) \f({\bf x}) \Biggr] 
\ee  
Here  $\t = (\T-T)/\T$ ($\t \ll 1$) is the reduced 
temperature parameter (for simplicity in what follows
it will be supposed that $\T =1$). According to this definition, 
positive values of $\t$ correspond to the low-temperature 
ferromagnetic state. Random fields $h({\bf x})$ are described by the 
symmetric Gaussian distribution,
\be
\lb{rfim2}
P[h({\bf x})] = p_{0} 
\exp \Biggl( -\frac{1}{2h^{2}}\I \; h^{2}({\bf x}) \Biggr) \ ,
\ee
where $h$ is the parameter which describes the strength of the random field, 
and $p_{0}$ is an irrelevant normalization constant. 
For a given realization of the random fields 
the partition function of this system is obtained by the integration over 
all configurations of the scalar fields $\f({\bf x})$

\be
\lb{rfim3}
Z[h({\bf x})] \; = \; \int {\cal D} \f({\bf x}) \exp\Biggl( -H\[\f({\bf x}),h({\bf x})\]\Biggr)
\ee
It is well known that if we consider such system at temperatures not too close to the critical point
(so that the parameter $\t$ is not too small) the thermal critical fluctuations are
irrelevant, and the leading contributions to the partition function, eq.(\ref{rfim3}),
are coming from the minima of the Hamiltonian, eq.(\ref{rfim1}), described by the 
saddle-point equation

\be
\lb{rfim4}
-\D\f({\bf x}) - \t \f({\bf x}) + g\f^{3}({\bf x}) \; = \; h({\bf x})
\ee  
In the pure system (at $h({\bf x}) = 0$) the restriction on the value of the parameter $\t$ 
is given by the Ginzburg-Landau condition

\be
\lb{rfim5}
\t \; \gg \; g^ {2/(4-D)} \; \equiv \t_{GL}
\ee
It is clear that this requirement makes sense only if the coupling
parameter $g$ is sufficiently small. Of course, formally for the three-dimensional 
Hamiltonian, eq.(\ref{rfim1}), considered as the continuous limit representation 
of the original Ising model this is not true (unlike the corresponding systems in dimensions 
close to four, where the effective (renormalized) value of the coupling parameter
$g \sim \epsilon = (4-D) \ll 1$). Nevertheless, here we are going to consider 
the random field Ginzburg-Landau theory, eq.(\ref{rfim1})
in which the coupling parameter $g$ is assumed to be sufficiently small,
hoping that at the qualitative level the behavior of the system is not so much
sensitive to the actual value of this parameter.

The only relevant spatial scale in the system described by
the Hamiltonian (\ref{rfim1}) is the correlation length $R_{c}(\t)$ which 
under condition (\ref{rfim5}) has the scaling

\be
\lb{rfim6}
R_{c}(\t) \; \sim \; \t^{-1/2}
\ee
In the absence of the random fields the ferromagnetic ground state of 
the system is characterized by the homogeneous configuration
$\f_{0} = \sqrt{\t/g}$. To study the effects produced by the random fields
on the ferromagnetic state of the system let us perform the following
space and fields rescaling

\be
\lb{rfim7}
\f({\bf x}) \; = \; \(\fr{\t}{g}\)^{1/2} \p({\bf x}/R_{c}) ,
\ee
In terms of the new fields $\p({\bf z})$, 
where ${\bf z} \equiv {\bf x}/R_{c} = \t^{1/2} {\bf x}$,
the system is described by the rescaled Hamiltonian

\be
\lb{rfim8}
H\[\p({\bf z}), \tl{h}({\bf z})\] \; = \; \fr{\t^{\fr{(4-D)}{2}}}{g} \int d^{D} z 
\Biggl[ \2 \(\n\p({\bf z})\)^{2}  
   - \2 \p^{2}({\bf z})  
   + \4 \p^{4}({\bf z}) -\sqrt{\fr{g}{\t^{3}}} \tl{h}({\bf z}) \p({\bf z}) \Biggr]
\ee 
where the rescaled random fields

\be
\lb{rfim9}
\tl{h} ({\bf x}/R_{c}) \; = \; 
R_{c}^{-D} \int_{|{\bf x}'-{\bf x}|<R_{c}} d^{D} x' \; h({\bf x}')
\ee
are described by the distribution function

\be
\lb{rfim10}
P[\tl{h}({\bf z})] = \tl{p}_{0} 
\exp\Biggl( -\frac{1}{2h^{2} \t^{D/2}}
         \int d^{D} z \tl{h}^{2}({\bf z}) \Biggr) \ ,
\ee
(note that the integration here involves the ultraviolet 
cutoff lengthscale equal to one). Redefining the random fields again

\be
\lb{rfim11}
\tl{h}({\bf z}) = \sqrt{\fr{\t^3}{g}} \xi({\bf z})
\ee
instead of eq.(\ref{rfim3}) we obtain  the partition function

\be
\lb{rfim12}
Z[\xi({\bf z})] \; = \; 
\int {\cal D} \p({\bf z}) \exp\Biggl( -\tl{\b} \tl{H}\[\p({\bf z}),\xi({\bf z})\]\Biggr)
\ee
which is controlled by the effective "inverse temperature"

\be
\lb{rfim13}
\tl{\b} \equiv \(\fr{\t}{\t_{GL}}\)^{\fr{(4-D)}{2}} \; \gg \; 1
\ee
($\t_{GL}$ is the Ginzburg-Landau temperature, eq(\ref{rfim5}))
and which is defined by the new effective Hamiltonian

\be
\lb{rfim14}
H\[\p({\bf z}), \xi({\bf z})\] \; = \;  \int d^{D} z \Biggl[ \2 \(\n\p({\bf z})\)^{2}  
   - \2 \p^{2}({\bf z})  + \4 \p^{4}({\bf z}) - \xi({\bf z}) \p({\bf z}) \Biggr]
\ee
which contains no parameters. The new random fields $\xi({\bf z})$ are described by 
the Gaussian distribution

\be
\lb{rfim15}
P[\xi({\bf z})] = p_{0} 
\exp \Biggl( -\frac{1}{2\lm^{2}}\I \; \xi^{2}({\bf z}) \Biggr) \ ,
\ee
characterized by the mean square value

\be
\lb{rfim16}
 \lm^2 \; = \;  \fr{(g h^2)}{ \t^{(6-D)/2}}
\ee
The ground state configurations in terms of the new fields $\p({\bf z})$
are defined by the saddle-point equation

\be
\lb{rfim17}
-\D\p({\bf z}) - \p({\bf z}) + \p^{3}({\bf z}) \; = \; \xi({\bf z})
\ee  

In terms of the new fields, $\p({\bf z})$ and $\xi({\bf z})$, the transition from 
the ordered (ferromagnetic) to the disordered (paramagnetic) state looks as follows.
In the absence of the random fields ($\xi \equiv 0$) the ground state is
given by  the trivial ferromagnetic solution $\p_{0}({\bf z}) = 1$ (or $\p_{0}({\bf z}) = -1$).
The presence of weak random fields (at  $\lm \ll 1$) introduces only small perturbations to this solution.
However, if we increase the effective strength of $\xi({\bf z})$, which is controlled by 
the parameter  $\lm$, eq.(\ref{rfim16}), the ferromagnetic configuration is getting more 
and more perturbed, and finally, at at a certain critical value $\lm_{c}$ the ferromagnetic 
ordering is destroyed.

One can note two important points in the above scenario:

{\bf First}. According eq.(\ref{rfim16}), for fixed value of the parameter $h$ 
(which is the strength of the original random fields $h({\bf x})$),
the effective strength of the random fields $\xi({\bf z})$ is controlled by the temperature
parameter $\t$. According to its definition ($\t = (1-T)$), increasing the temperature $T$ of the system 
means decreasing $\t$, which in turn, produces increasing of the parameter $\lm$
In other words, variations in temperature turn into variations of the effective strength
of the random fields.

{\bf Second}. According to the saddle-point equation (\ref{rfim17}) the critical value of $\lm_{c}$
is of the order of one. This means that the transition takes place at 

\be
\lb{rfim18}
\t \; \sim \; \t_{c} \; = \;  (g h^2)^{\fr{2}{(6-D)}}
\ee 
 At this point the value of the effective "inverse temperature" parameter $\tl{\b}$,
 eq.(\ref{rfim13}) is 

\be
\lb{rfim19}
\tl{\b_{c}} \; = \;  \(\fr{h^{2}}{g^{2/(4-D)}}\)^{\fr{(4-D)}{(6-D)}}
\ee
Thus,   if the strength of random fields is not too small:

\be
\lb{rfim20}
 h \; \gg \; h_{*}(g) \; = \; g^{\fr{1}{(4-D)}}
 \ee
 at dimensions $D < 4$ in the vicinity of the phase transition (when $\t \sim \t_{c}$) we have
 
\be
\lb{rfim21}
\tl{\b} \; \sim \; \tl{\b_{c}} \; = \;  \(\fr{{h}}{h_{*}(g)}\)^{\fr{2(4-D)}{(6-D)}} \; \gg \; 1
\ee
In other words, in terms of the representation, eqs.(\ref{rfim12})-(\ref{rfim16}), 
the original ferromagnetic-paramagnetic phase transition at any point
of the critical curve $h_{c}(T)$ on the left of the point ($h_{*}, T_{*}$) (Figure 1)
is equivalent to the low-temperature order-disorder transition (far left extreme of the curve $h_{c}(T)$)
at variation of the strength of the random fields.
In particular, in the dimension $D = 3$,

\be
\ba
\lb{rfim22}
\t_{GL} = g^2
\\
\\
\t_{c} \; = \;  (g h^2)^{2/3}
\\
\\
\tl{\b_{c}} \; = \;  \(\fr{h}{g}\)^{2/3}
\\
\\
 h_{*}(g) \; = \; g
 \ea
 \ee
The conditions $\t_{c} \gg \t_{GL}$, $ \tl{\b_{c}} \gg 1$ and $h \gg h_{*}$ discussed above are automatically satisfied by the only restriction on the strength of the random fields,

\be
\lb{rfim26}
 h \; \gg \; g
 \ee
In the next section to understand the nature of the low temperature phase transition,
we are going to consider its extreme version in the {\it zero temperature} limit. 
In this case it is natural to consider the original
discrete Ising model on a lattice instead of its continuous limit representation.

 \begin{figure}[h]
\begin{center}
\includegraphics[width=9cm]{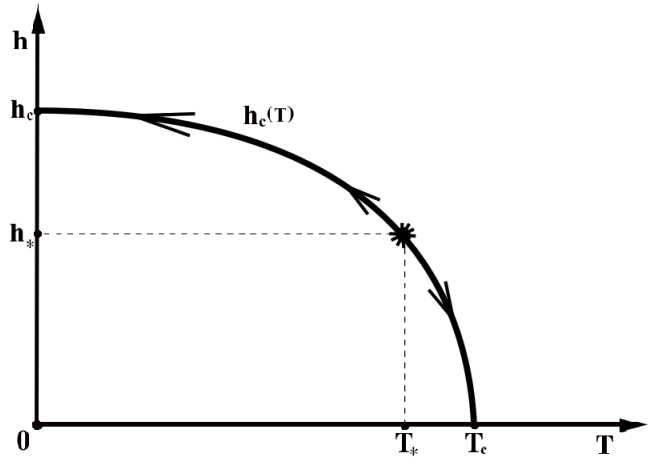}
\caption[]{Qualitative phase diagram of the random field Ising model}
  \label{fig1}
\end{center}
\end{figure}

\section{Zero-temperature phase transition}

Let us consider the three-dimensional random field ferromagnetic Ising model on the cubic
lattice with the Hamiltonian

\be
\lb{rfim27}
H \; = \; -\fr{1}{12} \sum_{<j\not= j>}^{N} \s_{i} \s_{j} \; - \; \sum_{i}^{N} h_{i} \s_{i}
\ee
Here $<j\not= j>$ denotes the pairs of nearest neighbors, and the random fields $h_{i}$ 
are described by the independent Gaussian distributions

\be
\lb{rfim28}
W_{h} [h_{1}, h_{2}, ..., h_{N}] \; = \; \prod_{i=1}^{N} \; {\cal P}_{h}(h_{i})
\ee
where

\be
\lb{rfim29}
{\cal P}_{h}(h_{i}) \; = \; \fr{1}{ \sqrt{2 \pi h^{2}}} \; \exp\(-\fr{h_{i}^{2}}{2 h^{2}}\)
\ee
The ground state spin configuration of this system is defined by the set of $N$ the conditions

\be
\lb{rfim30}
\s_{i} \; = \; Sign \( \fr{1}{6} \sum_{\a_{i} =1}^{6} \; \s_{\a_{i}} \; + \; h_{i} \)
\ee
where for every site $i$ the summation goes over its 6 nearest neighbors.

First, let us discuss at the qualitative level what is going on if we increase the parameter
$h$  from zero to large values. 
At $h=0$ the state of the system is ferromagnetic: the solution of the above equations (\ref{rfim30})
is trivial, $\s_{i} = +1$ (of course, there is another equivalent solution $\s_{i} = -1$, but in what follows 
we will suppose
that in the ferromagnetic state the spins are directed "up").   At non-zero $h \ll 1$ almost all the spins
are directed "up", but in rare cases, when at a given site $i$ the value of the local field is sufficiently
negative, $h_{i} < -1$, the $i$-th spin is flipped "down". According to eq.(\ref{rfim29}), 
the probability of this event (which is equal to the concentration  of the negative spins) is 

\be
\lb{rfim31}
\rho(h \ll 1) \; = \; \int_{-\infty}^{-1} dx \; {\cal P}_{h} (x) \; = \; \2 Erfc\(\fr{1}{\sqrt{2} h}\)
\ee
where

\be
\lb{rfim32}
Erfc(z) \; \equiv \; \fr{2}{\sqrt{\pi}} \int_{z}^{+\infty} \; dt \; \exp\(-t^{2}\)
\ee
is the complementary error function (which exponentially goes to zero at $z \gg 1$). 

Note that in the above consideration we have neglected the configurations in which the flipped down spins 
appear to be the nearest neighbors. It is clear  that at small $h$ the concentration $\rho(h)$ is exponentially 
small and therefore the probability of these events is negligible. However, when increasing the value 
of the parameter $h$ the probability to find two or more neighboring "down" spins becomes non-small, and
it is due to these configuration that the situation becomes rather sophisticated.
First of all, one can note that if at neighboring sites the values of the random fields appear to be
non small (and negative), then the stable spin configurations can become ambiguous. 
As an example let us consider two neighboring
sites $i$ and $i+1$ (surrounded by the spins which all are directed "up") 
with the values of the random fields 
$-1 < h_{i} < - 2/3$ and $-1 < h_{i+1} < - 2/3$. One can easily see that in this case both the configuration
$\s_{i} = \s_{i+1} = +1$ and the configuration $\s_{i} = \s_{i+1} = -1$ satisfy the stability conditions
eq.(\ref{rfim30}). 
It is obvious that one could  find similar phenomena in clusters consisting of larger number of spins.

The effects of the ambiguity of the stationary spin configurations due to the presence of the random fields
and their  (non-perturbative)  contributions to the finite-temperature 
thermodynamics have been discussed earlier in terms of the continuous Ginzburg-Landau representation 
of the RFIM \cite{3d-non-pert}. Here we will describe the consequences of the 
ambiguity of the stationary states for the phase transition at the zero temperatures. 
At the qualitative level it is tempting to suggest that the presence of such
kind of phenomena could be the origin of the phase transition of the first order. 
Indeed, if at increasing the strength of the random field  the system gets stuck at ferromagnetic 
configurations, an {\it alternative} disordered states could become  energetically preferable before the ferromagnetic 
state would become unstable. Note that unlike the second-order phase transitions which are characterized by
the divergence of the correlation length (and which require to find the way to study the system at large scales)
the first-order phase transition is characterized by a finite value of the correlation length. For that reason
one can hope that taking into account correlations only of the order of the lattice spacing could still
give reasonable results.

\subsection{Mean-field approach}
 
First, as a matter of simple "warming up exercise", let us consider how the the zero-temperature 
phase transition  looks like if we study it in terms of the ordinary mean-field approach 
(which neglects the spin-spin correlations at all).

\begin{figure}[h]
\begin{center}
\includegraphics[width=9cm]{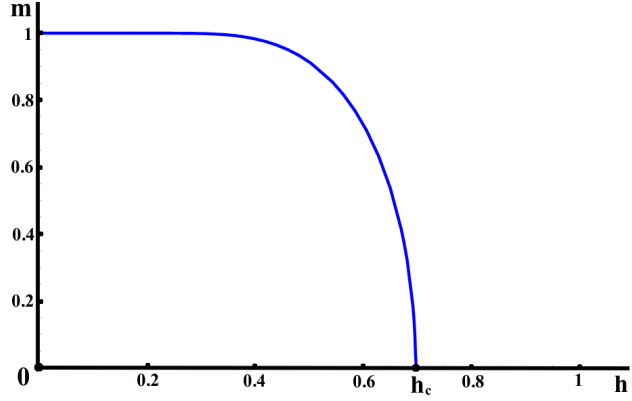}
\caption[]{Mean field solution for the order ferromagnetic order parameter $m(h)$ ($h_{c} \simeq 0.696$)}
  \label{fig2}
\end{center}  
\end{figure}

Let us denote by $x$ the probability that a given spin $i$ points "up". Then the probability to be "down"
is or course equal to $(1-x)$. In this case the ferromagnetic order parameter which is the global magnetization
of the system is 

\be
\lb{rfim35}
m \; = \; \fr{1}{N} \sum_{i}^{N} \s_{i} \; = \; \ol{\s_{i}} \; = \; 2 x - 1
\ee
A given spin is surrounded by the six nearest neighbors. 
If all the neighbors are pointing "up" (the probability of this configuration is equal to $x^{6}$)
then $\s_{i} = +1$  provided the local field $h_{i} > -1$. If one of the neighbors is
"down" (the probability is $6 x^{5} (1-x)$), then $\s_{i} = +1$  provided the local field $h_{i} > -2/3$.
If two of the neighbors are "down" (the probability is $\fr{6\cdot 5}{2}  x^{4} (1-x)^2$), then $\s_{i} = +1$  
provided the local field $h_{i} > -1/3$. Etc. Collecting all these configurations one easily gets
the following self-consistent equation for $x$:

\bq
\lb{rfim33}
x \; &=& \; x^{6} \cdot P_{6}(h) \; + \; 6 x^{5} (1-x) \cdot P_{5}(h) \; + \; 15 x^{4} (1-x)^{2} \cdot P_{4} (h) 
 \; + \; 20 x^{3} (1-x)^{3} \cdot P_{3}(h)  \; + \;
\nn
       &+& 15 x^{2} (1-x)^{4} \cdot P_{2} (h) \; + \; 6 x  (1-x)^{5} \cdot P_{1} (h) \; + \;  (1-x)^{6} \cdot P_{0} (h)
\eq
Here we have introduced the notations

\bq
\lb{rfim34}
P_{6}(h) &=& \int_{-1}^{\infty} dy {\cal P}_{h}(y) \; = \; 1 - \2 Erfc\(\fr{1}{\sqrt{2} h}\)
\nn
\nn
P_{5}(h) &=& \int_{-2/3}^{\infty} dy {\cal P}_{h}(y) \; = \; 1 - \2 Erfc\(\fr{2}{3\sqrt{2} h}\)
\nn
\nn
P_{4}(h) &=& \int_{-1/3}^{\infty} dy {\cal P}_{h}(y) \; = \; 1 - \2 Erfc\(\fr{1}{3\sqrt{2} h}\)
\nn
\nn
P_{3}(h) &=& \int_{0}^{\infty} dy {\cal P}_{h}(y) \; = \; \2
\nn
\nn
P_{2}(h) &=& \int_{1/3}^{\infty} dy {\cal P}_{h}(y) \; = \; \2 Erfc\(\fr{1}{3\sqrt{2} h}\)
\nn
\nn
P_{1}(h) &=& \int_{2/3}^{\infty} dy {\cal P}_{h}(y) \; = \; \2 Erfc\(\fr{2}{3\sqrt{2} h}\)
\nn
\nn
P_{0}(h) &=& \int_{1}^{\infty} dy {\cal P}_{h}(y) \; = \; \2 Erfc\(\fr{1}{\sqrt{2} h}\)
\eq

\begin{figure}[h]
\begin{center}
\includegraphics[width=10cm]{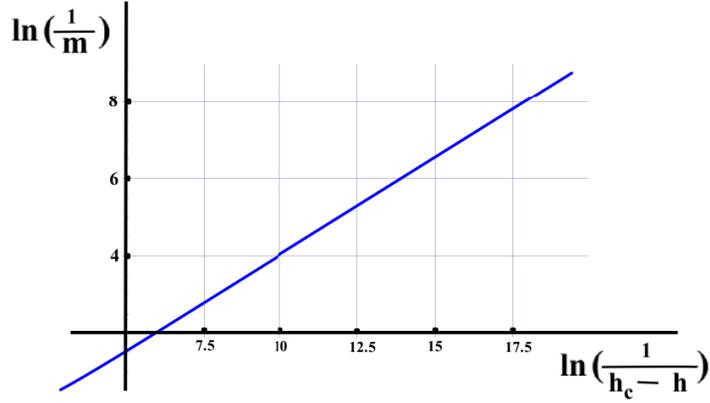}
\caption[]{Log-log plot for the ferromagnetic order parameter in the vicinity of the critical point}
  \label{fig3}
\end{center}  
\end{figure}

The solution of eq.(\ref{rfim33})  for the ferromagnetic order parameter is shown in Figure 2. We see that
the phase transition from the ferromagnetic ($m \not=  0$)  to the paramagnetic ($m=0$) state 
takes place at $h_{c} \simeq 0.696$.
One can easily check that in the vicinity of the critical point, as $h \to h_{c}$,  the the ferromagnetic 
order parameter vanishes according to the  the scaling law:
\be
\lb{rfim36}
m(h) \; \sim \; (h_{c} - h)^{\b}
\ee
described by the usual mean-field critical exponent $\b = 0.5$ (see Figure 3). 
This, of course, is not surprising because if in eq.(\ref{rfim33}) we substitute $x =(m+1)/2$,
the resulting equation  for $m$ (whatever complicated it may look like)
will be still the {\it mean field equation} for the order parameter, and its development
in powers of small $m(h)$ in the vicinity of the critical point can not give anything else but 
mean-field critical exponent.

It is clear that the consideration presented above can not pretend to describe the true phase
transition which takes place in the system under consideration. The reason for that is
obvious: in such king of the mean-field calculations one misses all spin-spin correlations,
which, as we have discussed in the beginning of this section, are crucial for the statistical
properties of the present system. 

Below we are going to modify the above approach so that it would  take into
account correlations between pairs of nearest neighbors spins. It turns out that
even such rather limited improvement is sufficient to intoduce rather dramatic changes in the
scenario of the phase transition.

\begin{figure}[h]
\begin{center}
\includegraphics[width=9cm]{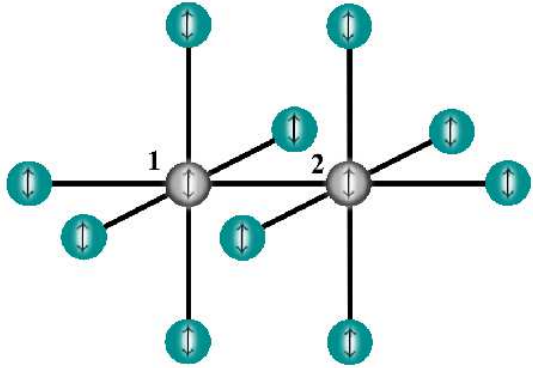}
\caption[]{Two neighboring sites are surrounded by (5+5) nearest neighbors}
  \label{fig4}
\end{center}
\end{figure}

\subsection{Modified mean-field approach}

Let us consider a pair of nearest neighbor spins  $\s_{1}$ and $\s_{2}$
and let us denote by $p(\uparrow\uparrow) $, $p(\uparrow\downarrow) $, $p(\downarrow\uparrow) $
and $p(\downarrow\downarrow) $ the probabilities of the $(\s_{1},\s_{2})$ configurations
$(+1, +1)$, $(+1, -1)$, $(-1, +1)$ and $(-1, -1)$ correspondingly. These four probabilities are,
of course,  bounded by the condition 
$p(\uparrow\uparrow)+p(\uparrow\downarrow)+p(\downarrow\uparrow) +p(\downarrow\downarrow) =1$.
As before, by $x$ we denote the probability for a given spin to be "up". One can easily see that

\be
\lb{rfim37}
x \; = \; p(\uparrow\uparrow) \; + \; p(\uparrow\downarrow)
\ee
The two spins $\s_{1}$ and $\s_{2}$ are surrounded by (5+5) neighbors (Figure 4), 
and their orientations are defined by the equations:

\bq
\lb{rfim38}
\s_{1} &=&  Sign \[ \fr{1}{6} \(\sum_{\a_{1} =1}^{5} \; \s_{\a_{1}} \; + \; \s_{2}\)  \; + \; h_{1} \]
\nn
\nn
\s_{2} &=&  Sign \[ \fr{1}{6} \(\sum_{\a_{2} =1}^{5} \; \s_{\a_{2}} \; + \; \s_{1}\)  \; + \; h_{2} \]
\eq
where $ \s_{\a_{1,2}}$ denote the neighbors of the spins $\s_{1,2}$.

The idea is to compute $ p(\uparrow\uparrow)$ and $p(\uparrow\downarrow)$
by summing over all possible configurations of these neighboring spins
(with the corresponding probabilities defined by the parameter $x$) and 
integrating over local fields $h_{1}$ and $h_{2}$ (with the probability
distribution ${\cal P}_{h}(h_{1})  {\cal P}_{h}(h_{2})$). Then, substituting 
$ p(\uparrow\uparrow)$ and $p(\uparrow\downarrow)$ (which 
will be the functions of $x$ and $h$) into  eq.(\ref{rfim37}),
 we will get a self-consistent equation for  $x$. It turns out, however,
 that this program can not be implemented directly, because for
 every configuration of surrounding spins there are finite regions in the plane
 $(h_{1}, h_{2})$, where the orientations of spins $\s_{1}$ and $\s_{2}$
 are {\it not} uniquely defined by the given values of $h_{1}$ and $ h_{2}$.
  \begin{figure}[h]
\begin{center}
\includegraphics[width=10cm]{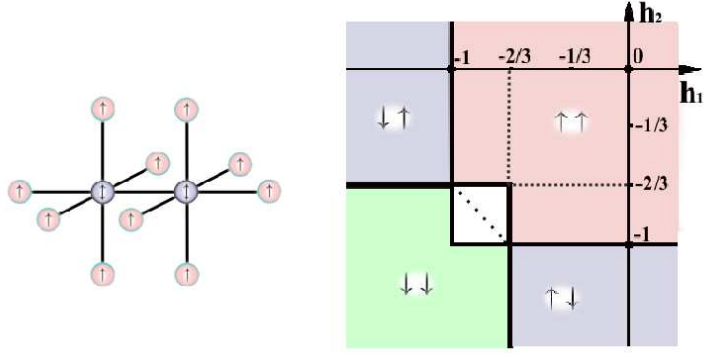}
\caption[]{The structure diagram of the orientations of two neighboring spins
for the case when all their (5+5) nearest neighbors are directed "up"}
  \label{fig5}
\end{center}
\end{figure}
\begin{figure}[h]
\begin{center}
\includegraphics[width=12cm]{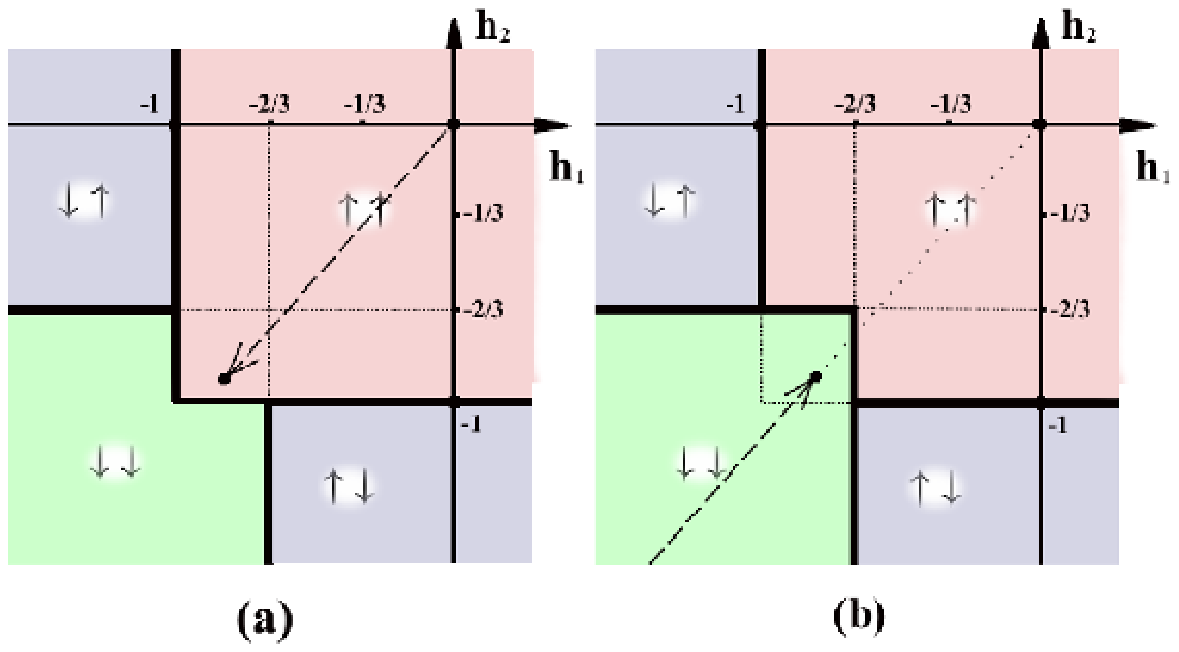}
\caption[]{The orientation diagram of two neighboring spins
for the case when all their (5+5) nearest neighbors are directed "up": {\bf (a)} at {\it increasing} and 
{\bf (b)} at {\it decreasing} strength $h$ of the random fields}
  \label{fig6}
\end{center}
\end{figure}
\begin{figure}[t]
\begin{center}
\includegraphics[width=11cm]{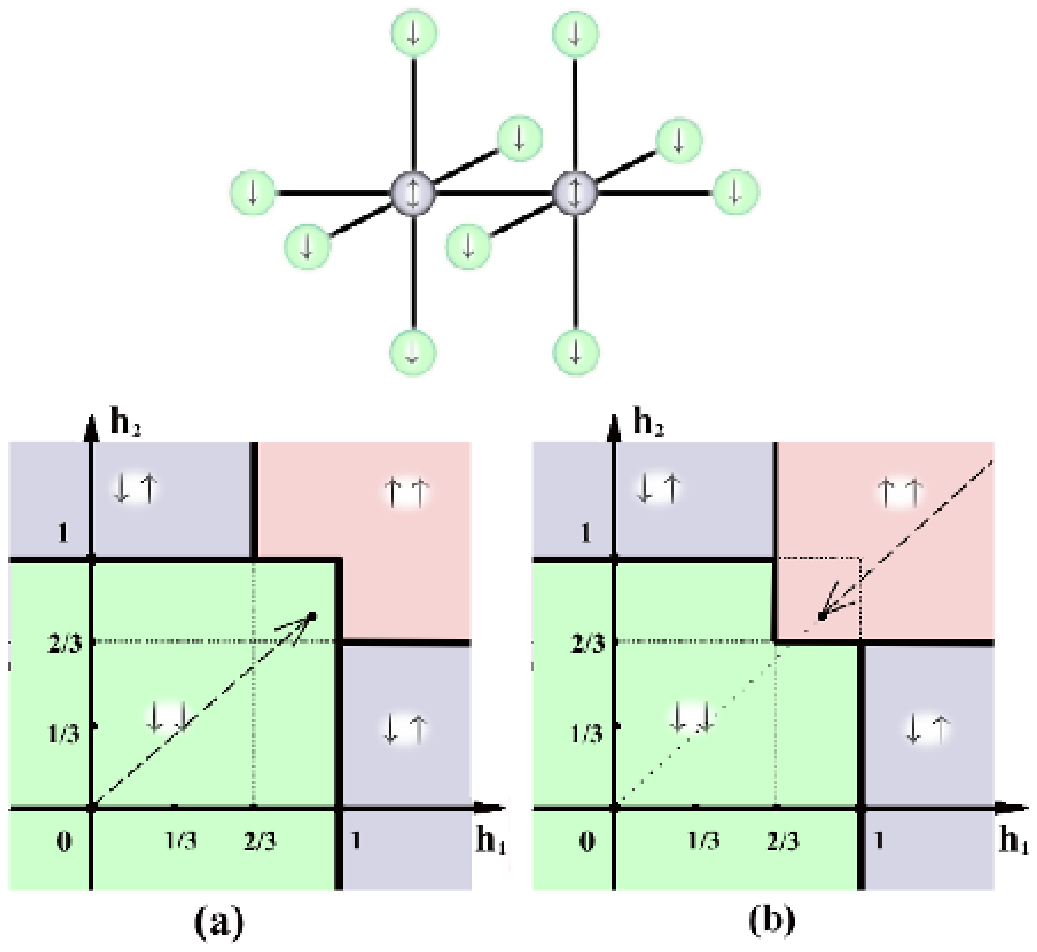}
\caption[]{The orientation diagram of two neighboring spins
for the case when all their (5+5) nearest neighbors are directed "down":  {\bf (a)} at increasing and 
{\bf (b)} at decreasing strength $h$ of the random fields}
  \label{fig7}
\end{center}
\end{figure} 
\begin{figure}
\begin{center}
\includegraphics[width=12cm]{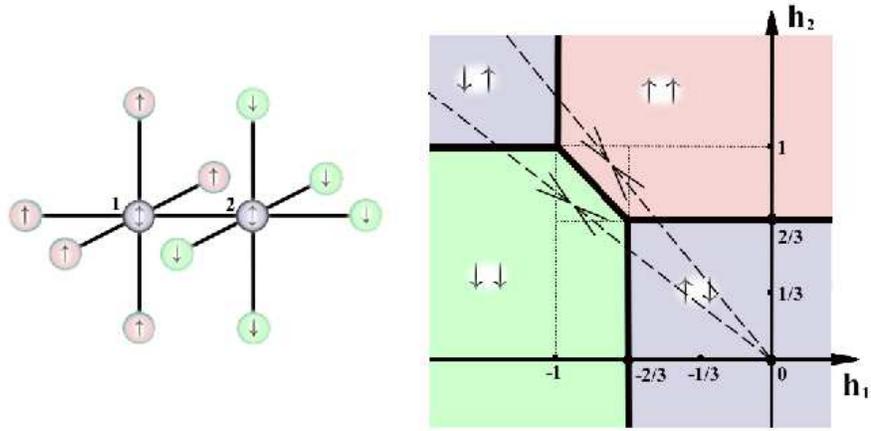}
\caption[]{The orientation diagram of two neighboring spins
for the case when all 5 neighbors of the spin $\s_{1}$ are directed "up", 
while all 5 neighbors of the spin $\s_{2}$ are directed "down"}
  \label{fig8}
\end{center}
\end{figure}
\begin{figure}
\begin{center}
\includegraphics[width=12cm]{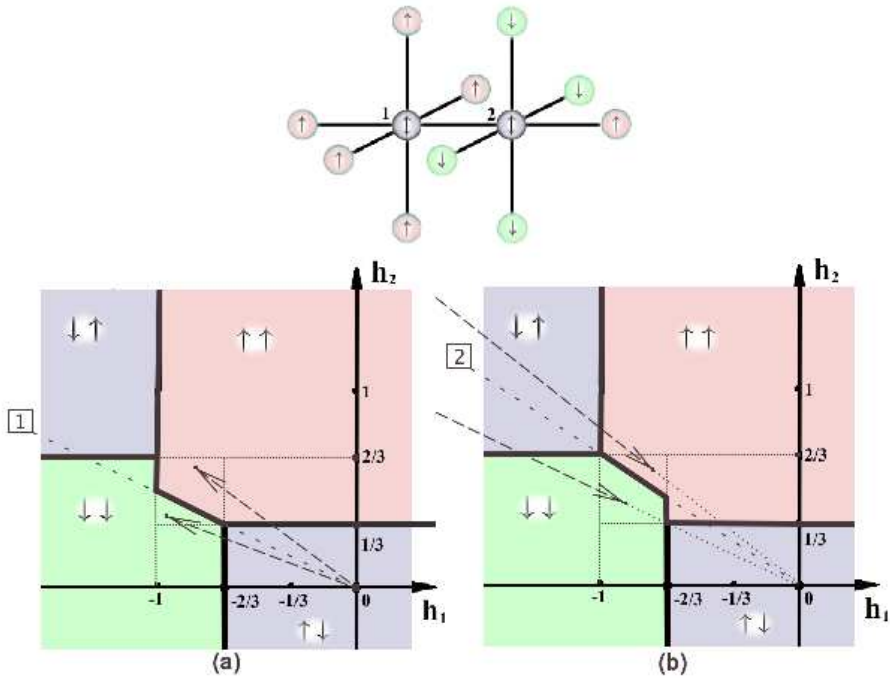}
\caption[]{The orientation diagram of two neighboring spins
for the case when all 5 neighbors of the spin $\s_{1}$ are directed "up", 
and 4  neighbors of the spin $\s_{2}$ are directed "down": {\bf (a)} at {\it increasing} and 
{\bf (b)} at {\it decreasing} strength $h$ of the random fields. 
The lines are:
[1] $h_{2}(h_{1}) = -(1/2) h_{1}$;  [2] $h_{2}(h_{1}) = -(2/3) h_{1}$}
  \label{fig9}
\end{center}
\end{figure}
\begin{figure}
\begin{center}
\includegraphics[width=10cm]{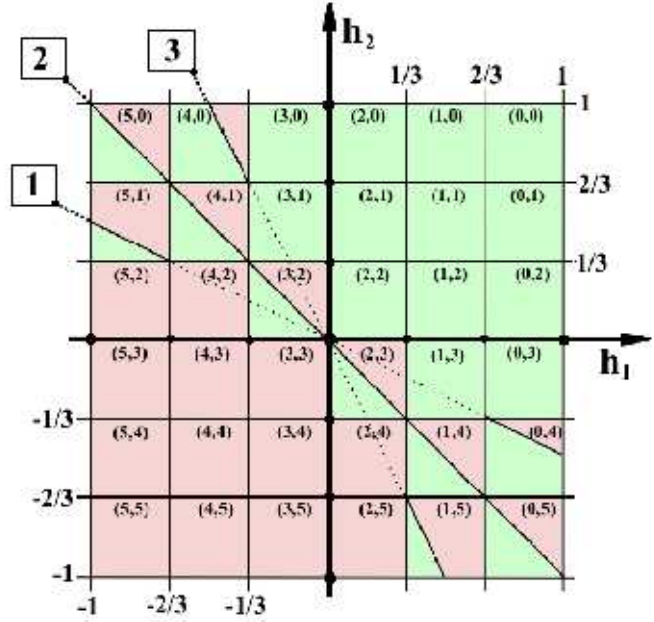}
\caption[]{The structure of the "frustrated squares" at {\it increasing} strength $h$
of the random fields. The notation {\bf $(m,n)$} inside the square indicate that 
it corresponds to the configuration with $m$ neighbors of the spin $\s_{1}$ 
and $n$ neighbors of the spin $\s_{2}$ are directed "up". Rose regions corresponds to
the state with both spins $\s_{1}$ and $\s_{2}$ are directed "up". Green regions are
the ones where both spins $\s_{1}$ and $\s_{2}$ are directed "down". The lines are:
[1] $h_{2}(h_{1}) = -(1/2) h_{1}$; [2] $h_{2}(h_{1}) = -h_{1}$; [3] $h_{2}(h_{1}) = -2 h_{1}$.} 
  \label{fig10}
\end{center}
\end{figure}
\begin{figure}
\begin{center}
\includegraphics[width=10cm]{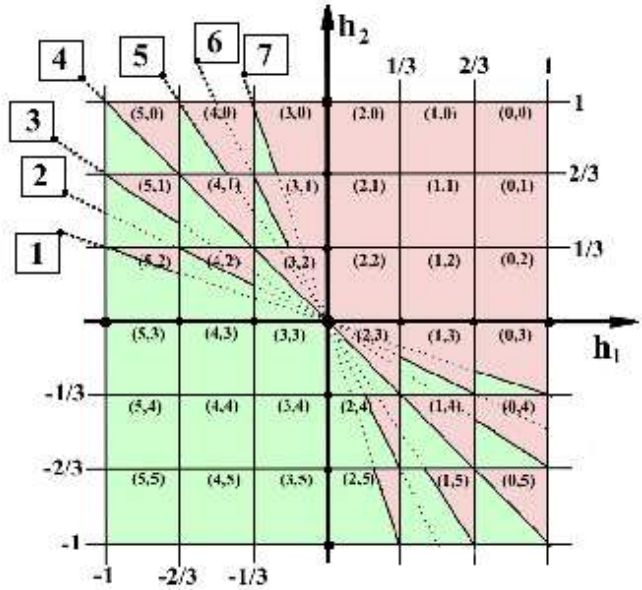}
\caption[]{The structure of the "frustrated squares" at {\it decreasing} strength $h$
of the random fields. Notations are the same as in Figure 10. The lines are:
[1] $h_{2}(h_{1}) = -(1/3) h_{1}$; [2] $h_{2}(h_{1}) = -(1/2)h_{1}$; 
[3] $h_{2}(h_{1}) = -(2/3) h_{1}$; [4] $h_{2}(h_{1}) = -h_{1}$; [5] $h_{2}(h_{1}) = -(3/2) h_{1}$;
[6] $h_{2}(h_{1}) = - 2 h_{1}$; [7] $h_{2}(h_{1}) = - 3 h_{1}$}
  \label{fig11}
\end{center}
\end{figure}
\begin{figure}[h]
\begin{center}
\includegraphics[width=11cm]{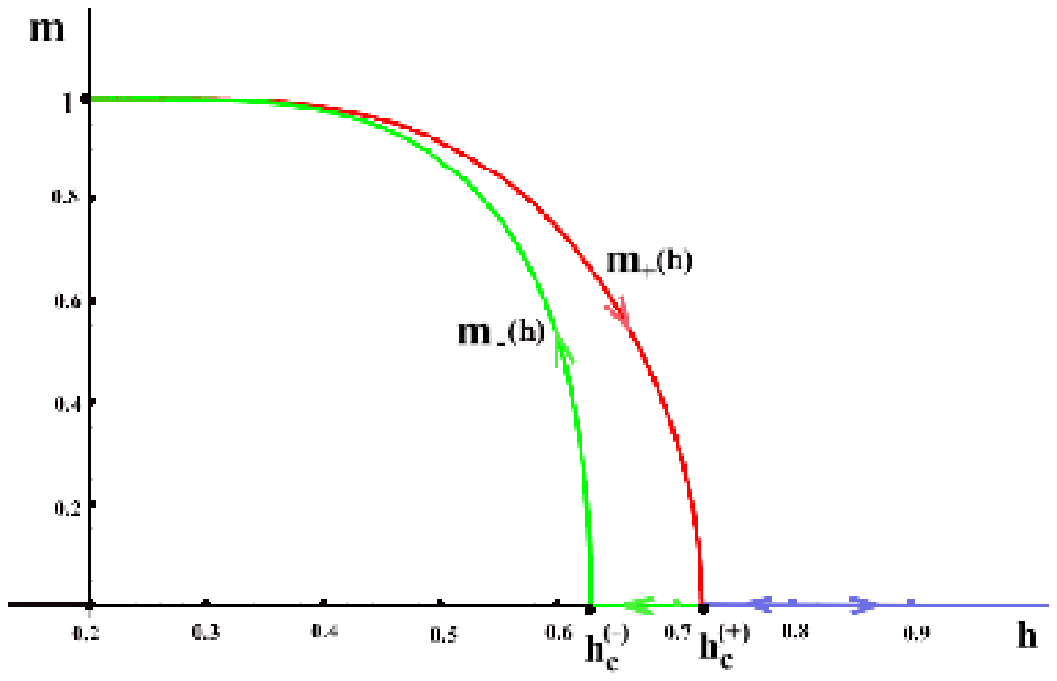}
\caption[]{Ferromagnetic order parameter $m(h)$ at  increasing (red line) and decreasing (green line)
strength $h$ of the random fields. $h_{c}^{(+)}\simeq 0.722; h_{c}^{(-)} \simeq 0.622$.}
  \label{fig12}
\end{center}
\end{figure}

 As an example let us consider again the simplest situation when all ten neighboring
spins are directed "up" (which has the probability $x^{10}$). In this case
the condition to have both spins"up" is $h_{1} > -1$ and $h_{2} > -1$ 
(so that the corresponding contribution to $p(\uparrow\uparrow)$ must be given by 
the integration of ${\cal P}_{h}(h_{1})  {\cal P}_{h}(h_{2})$ in the sector 
 $(h_{1} > -1, \; h_{2} > -1)$). The contribution to the 
probability $p(\uparrow\downarrow)$ is given by the sector  $(h_{1} > -2/3, \; h_{2} < -1)$. 
The contribution to the probability $p(\downarrow\uparrow)$ is given by the sector 
$(h_{1} < -1, \; h_{2} > -2/3)$. And finally the contribution to
the probability $ p(\downarrow\downarrow)$ is given by the sector 
 $(h_{1} < -2/3, \; h_{2} < -2/3)$, which {\it overlaps} with the sector corresponding
to  $p(\uparrow\uparrow)$. Thus in the square $(-1 < h_{1} < -2/3; \; -1 < h_{2} < -2/3)$
we are facing a kind of frustration (Figure 5). 

One can propose two ways out of this situation. First of all, one can, of course, compare the 
energies of the two configurations, and then, if the spin orientations always
correspond to the ground state, the ambiguity will be lifted (the configuration with both spins "up"
has lower energy in the triangle above the dotted line inside the "frustrated square" in Figure 5, and 
the one with both spins "down" has lower energy below the dotted line). 
However, if we keep in mind the scenario in which in the course of variations of the external
parameter $h$ each spin just follow the direction of the local field, while all
thermally activated "jumps over barriers" are suppressed, then we have to conclude that inside the 
"frustrated" region the orientation of the two spins must depend on the {\it history}.
Namely, for every point in the plane $(h_{1}, h_{2})$ we can assign the "trajectory"
which would demonstrate where this particular values of the random fields came from.
For instance, in the situation when we study the evolution of the system for {\it increasing} value of $h$, 
(starting from zero) a particular point $(h_{1}, h_{2})$ has the "trajectory" which is the straight line 
connecting this point with the origin. In the opposite case of {\it decreasing} $h$ (starting from infinity),
the "trajectory" of the point $(h_{1}, h_{2})$ is the straight line which comes from
infinity and directed towards the origin.

Following these rules of the game the ambiguity of the
spin orientations is lifted {\it provided} one fixes which process is actually under study:
increasing or decreasing of $h$.  In the study of the ferromagnetic $\to$ paramagnetic
transition (increasing $h$) one can easily note that the "trajectory" to any point inside
the "frustrated" square $(-1 < h_{1} < -2/3; \; -1 < h_{2} < -2/3)$ comes from the region
where both spins, $\s_{1}$ and $\s_{2}$, are directed "up". Thus, in this case
the square "belong" to the probability  $p(\uparrow\uparrow)$ (Figure 6(a)). 
In the reversed process of decreasing $h$ (the paramagnetic $\to$ ferromagnetic transition)
the only way to arrive inside this square is to come from the region where both spins are "down",
and in this case the square belong to the probability $ p(\downarrow\downarrow)$ (Figure 6(b)).

For better understanding how it works let us consider other examples. In the case when all 10
surrounding spins are directed "down", the "orientation diagram" for the spins $\s_{1}$ and $\s_{2}$
is represented in Figure 7. Here the "frustrated square" is located in the region
$(2/3 < h_{1} < 1; \; 2/3 < h_{2} < 1)$. Now, at increasing $h$ one arrives into this square from
the region where both spins are "down" and therefore in this case it belongs to the probability
$ p(\downarrow\downarrow)$ (Figure 7(a)). Apparently, in the reversed process of decreasing $h$ 
the square belong to the probability $p(\uparrow\uparrow)$ (Figure 7(b)).

Let us consider the configuration in which all 5 neighbors of the spin $\s_{1}$ are "up", 
while all 5 neighbors of the spin $\s_{2}$ are "down". Using eqs.(\ref{rfim38}) one can easily
build the corresponding orientation diagram (Figure 8). We see that in this case
both at increasing  and decreasing $h$ one can arrive into the left-down triangle of 
the "frustrated square" only by passing  the region where both spins are "down". 
Similarly, both at increasing and decreasing $h$  one can arrive into the right-up triangle 
only by passing  the region where both spins are "up".  

Finally, let us consider the configuration in which all  5 neighbors of the spin $\s_{1}$ are "up", 
and only 4  neighbors of the spin $\s_{2}$ are "down". Here the division of the "frustrated square"
between $p(\uparrow\uparrow)$ and $ p(\downarrow\downarrow)$ is somewhat more
complicated (Figure 9). At increasing $h$ one arrives into the section of the square below the
line $h_{2}(h_{1}) = -\fr{1}{2} h_{1}$ by passing through the region where both spins are "down"
and one arrives into the section above this line by passing through the region where both spins
are "up" (Figure 9(a)).  On the other hand, at decreasing $h$ one arrives into the section of the square 
below the line $h_{2}(h_{1}) = -\fr{2}{3} h_{1}$ by passing through the region where both spins 
are "down" and one arrives into the section above this line by passing through the region 
where both spins are "up" (Figure 9(b)).

Now, with some patience and perseverance one can construct the structure of the 
frustrated squares for all 36 "analytically different" (in terms of eqs.(\ref{rfim38}))
configurations of the surrounding spins.
The resulting orientation diagrams for increasing  and  decreasing $h$ are shown in 
Figures 10 and 11 correspondingly.

Finally, after the regions corresponding to "up-up" and "up-down" orientations
of the spins $\s_{1}$ and $\s_{2}$ are unambiguously defined, the derivation 
of the equations for the probabilities 
$ p(\uparrow\uparrow)(x, h)$ and $p(\uparrow\downarrow)(x,h)$ is straitforward
(although slightly cumbersome). The explicit forms of the equation (\ref{rfim37})
for increasing and for decreasing $h$ is given in the Appendix.

The solution of these equations in terms of the ferromagnetic order parameter $m(h)$
is represented in Figure 12.
We see that the behavior of the magnetization exhibits clear hysteresis phenomenon.

\section{Discussion}

In this paper the zero-temperature phase transition has been studied
under the assumption that at any variations of the strength $h$ of the random fields
the transformations of the spin configurations go under constrain that 
all thermally activated spin flips are suppressed. 
It should be stressed that this situation is essentially different from the 
true equilibrium phase transition where at any given $h$ 
the system is supposed to be in the ground state (and which, in my view,
is experimentally inaccessible, et least in the low-temperature limit).

Comparing two possible scenarios of the order-disorder phase transition:
continuous (the second order) and discontinuous (the first order), one 
should note that unlike the continuous transition characterized by the
divergence of the correlation length, at the first-order phase transition the 
correlation length remains finite. Thus, admitting that the transition is
discontinuous one can hope that a theory which takes into account
spin-spin correlations only at a limited scale (e.g. of the order of the
lattice spacing) still would give qualitatively correct description
of the phase transition.

Usual mean-field theory is useless here because it does not take into account
spin-spin correlations at all. On the other hand, it has been demonstarted 
in this paper that even rather limited improvement of the mean-field
approach, which takes into account two nearest neighbor spins correlations,
produces rather dramatic effect on the scenario of the phase transition. 
After this modification the value of the ferromagnetic order parameter $m(h)$ 
as the function of the strength of the random fields becomes 
{\it history dependent} exhibiting clear hysteresis phenomenon (Figure 12). 

Unfortunately in the framework of the present theory it is still difficult to
to make a definite conclusion about the nature of the phase transition. 
The existence of the hysteresis loop (together with the clear understanding
of the physical mechanism in its origin) is the strong argument in favor of 
the first-order phase transition. On the other hand, each of the curves
$m_{+}(h)$ (at increasing $h$) and $m_{-}(h)$ (at decreasing $h$) in the Figure 12
demonstrates the continuous transition which, of course, makes no sense,
because at such transition the correlation length diverges, while the present theory
takes into account the correlations only of the order of the lattice spacing. 
In this sense the present theory is not self-consistent, and it would be reasonable 
to expect that the presence of the continuous transitions at $h_{c}^{(\pm)}$
is not more than just an artifact of the proposed approach.
These issues require further detailed studies.
 
 \vspace{10mm}

{\bf \large APPENDIX}

\vspace{5mm}

Let us denote by $p_{+}(\uparrow\uparrow) $ and $p_{-}(\uparrow\uparrow) $
the probabilities for the two spins $\s_{1}$ and $\s_{2}$ to be both "up" at increasing 
and at decreasing variations of $h$ correspondingly. Note that the 
probability $p(\uparrow\downarrow) $ is the same in both (increasing and decreasing) cases.
Then, using the orientation diagrams shown in Figures 10 and 11
(after some work) we get:

\be
\ba
\lb{rfim39}
p_{+}(\uparrow\uparrow) (x,h) \; = \;  
P_{6}(h) P_{6}(h) x^{10} + 10 P_{6}(h) P_{5}(h) x^9 (1 - x)^1
\\
\\
+\[ 20 P_{6}(h) P_{4}(h) + 25 P_{5}(h) P_{5}(h)\] x^8 (1 - x)^2 
+\[ 20 P_{6}(h) P_{3}(h) + 100 P_{5}(h) P_{4}(h) \] x^7 (1 - x)^3
\\
\\
+\biggl[ 10 \[P_{6}(h) P_{2}(h) - D^{(+)}_{51}(h)\]  
                                 + 100 P_{5}(h) P_{3}(h) + 100 P_{4}(h) P_{4}(h) \biggr] x^6 (1 - x)^4 
\\
\\
+\biggl[ 2 \[P_{6}(h) P_{1}(h) - 0.5 (P_{6}(h) - P_{5}(h)) (P_{1}(h) - P_{0}(h)) \] + 
         50 \[P_{5}(h) P_{2}(h) - 0.5 (P_{5}(h) - P_{4}(h)) (P_{2}(h) - P_{1}(h)) \]
\\
\; \; \; \; \; \;   
+200 \[P_{4}(h) P_{3}(h) - 0.5 (P_{4}(h) - P_{3}(h)) (P_{3}(h) - P_{2}(h)) \] \biggr] x^5 (1 - x)^5 
\\
\\
+\biggl[ 10 \[P_{5}(h) P_{1}(h) - (P_{5}(h) - P_{4}(h)) (P_{1}(h) - P_{0}(h)) + D^{(+)}_{40}(h) \] + 
                 100 \[P_{4}(h) P_{2}(h) - (P_{4}(h) - P_{3}(h)) (P_{2}(h) - P_{1}(h)) \] 
\\
\; \; \; \; \; \; 
+100 \[P_{3}(h) P_{3}(h) - (P_{3}(h) - P_{2}(h)) (P_{3}(h) - P_{2}(h)) \] \biggr]  x^4 (1 - x)^6 
\\
\\
+\biggl[ 20 \[P_{4}(h) P_{1}(h) - (P_{4}(h) - P_{3}(h)) (P_{1}(h) - P_{0}(h)) \] + 
         100 \[P_{3}(h) P_{2}(h) - (P_{3}(h) - P_{2}(h)) (P_{2}(h) - P_{1}(h)) \] \biggr] x^3 (1 - x)^7 
\\
\\ 
+\biggl[ 20 \(P_{3}(h) P_{1}(h) - (P_{3}(h) - P_{2}(h)) (P_{1}(h) - P_{0}(h)) \) + 
           25 \(P_{2}(h) P_{2}(h) - (P_{2}(h) - P_{1}(h)) (P_{2}(h) - P_{1}(h)) \) \biggr] x^2 (1 - x)^8 
\\
\\
+10 \[ P_{2}(h) P_{1}(h) - (P_{2}(h) - P_{1}(h)) (P_{1}(h) - P_{0}(h)) \] x^1 (1 - x)^9 
\\
\\
+ \[ P_{1}(h) P_{1}(h) - (P_{1}(h) - P_{0}(h)) (P_{1}(h) - P_{0}(h)) \] (1 - x)^{10}
\ea
\ee

\be
\ba
\lb{rfim40}
p_{-}(\uparrow\uparrow)(x,h) \; = \;
\\
 = \biggl[P_{6}(h) P_{6}(h) - (P_{6}(h) - P_{5}(h)) (P_{6}(h) - P_{5}(h))\biggr] x^{10} 
+ 10 \biggl[P_{6}(h) P_{5}(h) - (P_{6}(h) - P_{5}(h)) (P_{5}(h) - P_{4}(h))\biggr] x^9 (1 - x)^1 
\\
+ \biggl[20 \[P_{6}(h) P_{4}(h) - (P_{6}(h) - P_{5}(h)) (P_{4}(h) - P_{3}(h))\] + 
             25 \[P_{5}(h) P_{5}(h) - (P_{5}(h) - P_{4}(h)) (P_{5}(h) - P_{4}(h))\] \biggr] x^8 (1 - x)^2 
\\
+ \biggl[20 \[P_{6}(h) P_{3}(h) - (P_{6}(h) - P_{5}(h)) (P_{3}(h) - P_{2}(h)) + D^{(-)}_{52}(h)\] 
\\
 \; \; \; \; \; \;\; \; \; \; \; \;
        + 100 \[P_{5}(h) P_{4}(h) - (P_{5}(h) - P_{4}(h)) (P_{4}(h) - P_{3}(h))\] \biggr]  x^7 (1 - x)^3 
\\
+ \biggl[10 \[P_{6}(h) P_{2}(h) - (P_{6}(h) - P_{5}(h)) (P_{2}(h) - P_{1}(h)) + D^{(-)}_{51}(h)\] 
\\
\; \; \; \; \; \;\; \; \; \; \; \;
        + 100 \[P_{5}(h) P_{3}(h) - (P_{5}(h) - P_{4}(h)) (P_{3}(h) - P_{2}(h)) + D^{(-)}_{42}(h)\]
\\ 
\; \; \; \; \; \;\; \; \; \; \; \;         
        +100 \[P_{4}(h) P_{4}(h) - (P_{4}(h) - P_{3}(h)) (P_{4}(h) - P_{3}(h))\]\biggr] x^6 (1 - x)^4 
\\
+ \biggl[2 \[P_{6}(h) P_{1}(h) - 0.5 (P_{6}(h) - P_{5}(h)) (P_{1}(h) - P_{0}(h))\] + 
           50 \[P_{5}(h) P_{2}(h) - 0.5 (P_{5}(h) - P_{4}(h)) (P_{2}(h) - P_{1}(h))\] 
\\
\; \; \; \; \; \; \; \; \; \; \; \;
       + 200 \[P_{4}(h) P_{3}(h) - 0.5 (P_{4}(h) - P_{3}(h)) (P_{3}(h) - P_{2}(h))\] \biggr] x^5 (1 - x)^5 
\\
+ \biggl[10 \[P_{5}(h) P_{1}(h) - D^{(-)}_{51}(h)\] 
        + 100 \[P_{4}(h) P_{2}(h) - D^{(-)}_{42}(h)\] + 100 P_{3}(h) P_{3}(h) \biggr] x^4 (1 - x)^6 
\\
+  \biggl[20 \[P_{4}(h) P_{1}(h) - D^{(-)}_{52}(h)\] + 100 P_{3}(h) P_{2}(h)\biggr] x^3 (1 - x)^7 
\\
  +\[20 P_{3}(h) P_{1}(h) + 25 P_{2}(h) P_{2}(h)\] x^2 (1 - x)^8 
    +10 P_{2}(h) P_{1}(h)  x^1 (1 - x)^9 + P_{1}(h) P_{1}(h) (1 - x)^{10}
\ea
\ee

\be
\ba
\lb{rfim41}
p(\uparrow\downarrow)(x,h) \; = \; 
P_{5}(h) P_{0}(h) x^{10} + \[ P_{5}(h) P_{1}(h) + 5 P_{4}(h) P_{0}(h)\]  x^9 (1 - x)^1 
\\
\\
+\[10 P_{5}(h) P_{2}(h) + 25 P_{4}(h) P_{1}(h) + 10 P_{3}(h) P_{0}(h)\] x^8 (1 - x)^2
\\
\\
+ \[10 P_{5}(h) P_{3}(h) + 50 P_{4}(h) P_{2}(h) + 50 P_{3}(h) P_{1}(h) + 10 P_{2}(h) P_{0}(h)\] x^7 (1 - x)^3 
\\
\\
+ \[5 P_{5}(h) P_{4}(h) + 50 P_{4}(h) P_{3}(h) + 100 P_{3}(h) P_{2}(h) + 
        50 P_{2}(h) P_{1}(h) + 5 P_{1}(h) P_{0}(h)\]  x^6 (1 - x)^4 
\\
\\
+ \[P_{5}(h) P_{5}(h) + 25 P_{4}(h) P_{4}(h) + 100 P_{3}(h) P_{3}(h) + 
        100 P_{2}(h) P_{2}(h) + 25 P_{1}(h) P_{1}(h) + P_{0}(h) P_{0}(h)\]  x^5 (1 - x)^5 
\\
\\
+ \[5 P_{4}(h) P_{5}(h) + 50 P_{3}(h) P_{4}(h) + 100 P_{2}(h) P_{3}(h) + 
        50 P_{1}(h) P_{2}(h) + 5 P_{0}(h) P_{1}(h)\]  x^4 (1 - x)^6 
\\
\\
+ \[10 P_{3}(h) P_{5}(h) + 50 P_{2}(h) P_{4}(h) + 50 P_{1}(h) P_{3}(h) + 
        10 P_{0}(h) P_{2}(h)\]  x^3 (1 - x)^7 
\\
\\
+ \[10 P_{2}(h) P_{5}(h) + 25 P_{1}(h) P_{4}(h) + 10 P_{0}(h) P_{3}(h)\]   x^2 (1 - x)^8 
\\
\\
+ \[5 P_{1}(h) P_{5}(h) + 5 P_{0}(h) P_{4}(h)\] x^1 (1 - x)^9 +  P_{0}(h) P_{5}(h) (1 - x)^{10}
\ea
\ee
where the functions $P_{k}(h)\; \; $ $(k = 0,...,6)$ are defined in eq.(\ref{rfim34})
and 

\bq
\lb{rfim42}
D^{(+)}_{51}(h) &=&  \int_{-1}^{-2/3} dy_{1} {\cal P}_{h}(y_{1}) 
                                    \int_{1/3}^{-y_{1}/2} dy_{2} {\cal P}_{h}(y_{2})
\nn
\nn
D^{(+)}_{40}(h) &=&  \int_{2/3}^{1} dy_{2} {\cal P}_{h}(y_{2}) 
                                    \int_{-y_{2}/2}^{-1/3} dy_{1} {\cal P}_{h}(y_{1})
\nn
\nn
D^{(-)}_{52}(h) &=&  \int_{-1}^{-2/3} dy_{1} {\cal P}_{h}(y_{1}) 
                                    \int_{-y_{1}/3}^{1/3} dy_{2} {\cal P}_{h}(y_{2})
\nn
\nn
D^{(-)}_{51}(h) &=&  \int_{-1}^{-2/3} dy_{1} {\cal P}_{h}(y_{1}) 
                                    \int_{-2y_{1}/3}^{2/3} dy_{2} {\cal P}_{h}(y_{2})
\nn
\nn
D^{(-)}_{42}(h) &=&  \int_{-2/3}^{-1/3} dy_{1} {\cal P}_{h}(y_{1}) 
                                    \int_{-y_{1}/2}^{1/3} dy_{2} {\cal P}_{h}(y_{2})
\eq 
The ferromagnetic order parameters $m_{\pm}(h)$ 
as the function of  $h$ are obtained from the relation

\be
\lb{rfim43}
m_{\pm}(h) \; = \; 2 x_{\pm}(h) \; - \; 1
\ee
where $x_{+}(h)$ and $x_{-}(h)$ are the corresponding solutions of the equations

\be
\lb{rfim44}
x \; = \; p_{\pm}(\uparrow\uparrow)(x,h) \; + \; p(\uparrow\downarrow)(x,h)
\ee

\end{document}